\documentclass[twocolumn,superscriptaddress,pre]{revtex4}

\usepackage{helvet}
\usepackage{graphicx}
\usepackage{amsfonts,amsmath,amssymb}
\usepackage[T1]{fontenc}
\usepackage{ae,aecompl}

\begin{document}

\title{Modelling traffic flow fluctuations}
\author{Peter Wagner}
\affiliation{Institute of Transport Research, German Aerospace Centre\\
Rutherfordstrasse 2, 12489 Berlin, Germany}
\date{\today}


\begin{abstract}
  By analyzing empirical time headway distributions of traffic flow, a
  hypothesis about the underlying stochastic process can be drawn. The
  results found lead to the assumption that the headways $T_i$ of
  individual vehicles follow a linear stochastic process with
  multiplicative noise, $\dot T_i = \alpha (m_T - T_i) + D \, T_i\xi$.
  The resulting stationary distribution has a power-law tail,
  especially for densities where cars are interacting strongly.
  Analyzing additionally the headways for accelerating and
  decelerating cars, the slow-to-start effect proposed as a mechanism
  for traffic jam stability can be demonstrated explicitly. Finally,
  the standard deviation of the speed differences between following
  cars can be used to get a clear characterization of (at least) three
  different regimes of traffic flow that can be identified in the
  data.  Using the empirical results to enhance a microscopic traffic
  flow model, it can be demonstrated that such a model describes the
  fluctuations of traffic flow quite satisfactorily.
\end{abstract}

\maketitle

\section{Introduction}

Car drivers keep a certain distance to the car ahead, the space
headway $g$. It is defined as the distance between the front bumper of
the following to the rear bumper of the leading car. Obviously, this
distance depends on the speed $v$ of the cars, the bigger the speed,
the bigger is the distance. Therefore, it is useful to define the
(net) time headway $T$ as the scaled distance:
\begin{equation}
T := g/v \label{eq:def-time-headway}
\end{equation}
Very simple traffic flow models assume, that $T$ is constant for a
given driver, or even constant for a given ensemble of
drivers. Although this is certainly not true, no ''law'' has been
verified so far for the dependence of $T$ on $v$.

Note, that the headway definition Eq.~(\ref{eq:def-time-headway}) is
slightly different from the one usually used in the literature: there,
the time headway $\widetilde T$ is defined as the time between two
cars which have passed the same detector. In case that the cars do not
accelerate, the two definitions coincide, otherwise a correction has
to be applied to transform $T$ into the headways that have been
measured by those detectors (let $x(t)$ be the distance of the leading
car to the detector):
\[
T  =  \frac{g}{v} = \frac{1}{v} \int_0^{\widetilde T} x(t) dt  \approx \frac{v
\widetilde T + a \widetilde T^2/2}{v}  =  \widetilde T + \frac{a}{2v}\widetilde
T^2
\]
Usually, $a$ is small, so $T \approx \widetilde T$ follows, except for
small $v$ and large $\widetilde T$.

In the following, the distribution of the headways $p(T)$ will be the
object to study instead of individual headways. It contains useful
information about the interaction between the cars, and it is mostly
responsible for the fluctuations observed in traffic flow, even on a
macroscopic scale. Consequently, the traffic engineering literature
has a long record of different assumptions to describe the empirically
observed headway distributions
\cite{Adams:1936,Cowan:1976,Luttinen:1992}. Usually, the underlying
process is not stated explicitly, except in the case of free flow,
where a Poissonian process is assumed. Clearly, three different
regimes of traffic can be observed, which have to be discussed
separately. For small speeds $v$, corresponding to a jam, the
empirical data are not sufficient to draw solid conclusions. For large
speeds, corresponding to free flow, the distribution should finally
approach a Poissonian distribution since the cars do not interact with
each other anymore. The interesting regime is in between, for speeds
in the range $10 < v < 30$ m/s, where cars are interacting heavily.

Usually, variants of the Poisson distribution are being used for
describing headway distributions. Two popular examples of this are the
shifted exponential function $p(T)\propto \exp((T-T_0)/m_T)$, or the
Erlang function $p(T) \propto T^k \exp(-T/m_T)$, where $k$ is an
integer. Also, the log-normal function $p(T) = \exp(-((\log(T) -
m_T)/\sigma)^2/2)/( \sqrt{2 \pi} \sigma T)$ had been proposed for the
headways.

More appealing from a physical point of view is the enterprize to
relate the observed time headway distributions to a one-dimensional
electron gas with repelling interactions
\cite{Krbalek2001,Krbalek2003}. The latter idea leads to a so
called gamma distribution of the time headways that is given as
follows:
\begin{equation}
p(T) = \frac{( T - T_0 )^{\gamma - 1}}{\Gamma ( \alpha ) m_T^{\gamma}}
 \exp \left ( - \frac{T - T_0 }{ m_T } \right)  \quad T > T_0
\end{equation}
%
with the parameters $T_0$ (minimum headway), $m_T$ (where $T_0+m_T$ is
the mean headway), and $\gamma$, the so called shape parameter.
(Traffic engineers name this distribution Pearson type III.)

All these approaches yield sensible results when compared to reality.
However, they are not completely convincing, which can be seen by the
fact that still no common agreement in the ''right'' formulation has
been emerged. Probably, this is due to a shortage in reliable and
statistically meaningful data. Fortunately, this is going to change,
and this work has greatly benefitted from the availability of more
detailed microscopic data. The analysis presented below allow for an
alternative formulation, which will be worked out in the
section~\ref{sec:uFund} and will be implemented in a certain
microscopic car-following model in section~\ref{sec:YAM}. By doing so,
conclusions can be drawn about the macroscopic description of traffic
flow.

\section{Microscopic fundamental diagram}\label{sec:uFund}

\begin{figure}[ht]
\begin{center}
\includegraphics[width=\linewidth]{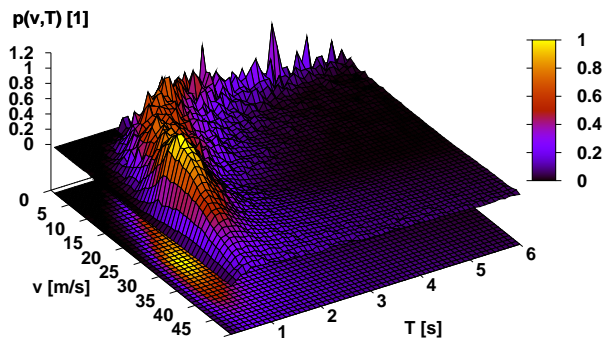}
\caption{Frequency distribution $p(v,T)$ of the net headways $T$ 
  as function of speed $v$. The raw data are from the left lane of the
  German autobahn A3 \cite{Schad:data:2002}. To determine the net
  headways, only cars with roughly the same lengthes have been used
  for this and in the subsequent analysis. This was necessary, since
  the data contain only the gross headway and the car lengthes.}
\label{fig:uFD-3D}
\end{center}
\end{figure}

The vast majority of all traffic flow data are collected by induction
loops at a certain place on the road. Therefore, just three things can
be measured directly and easily (if the detectors are double loop
detectors): the speed $v_i$ of each of the vehicles crossing the
detector, the length $l_i$ of this vehicle and the time headway $T_i$
between this and the preceding car. When using the net headway, the
uninteresting dependence on the car-lengths drops out. The plot of the
net headways $T_i$ as a function of the speeds $v_i$, or much better,
the corresponding distribution $p(v,T)$, will be called the
microscopic fundamental diagram in the following. An example is
presented in Fig.~\ref{fig:uFD-3D}, where $p(v,T)$ has been drawn for
the left lane of the German freeway A3.
Three regimes could be identified in Fig.~\ref{fig:uFD-3D}. A small
speed regime, where the mean headway seems to diverge. An intermediate
regime, where the distribution gets relatively small, and finally, the
high speed regime where the distribution and the mean headways get
fairly large.

The conventional macroscopic fundamental diagram can be obtained from
these data by plotting short-time averages of the speed $\langle v
\rangle $ versus flow $q$, where the latter is computed as $q =
1/\langle T \rangle$ (with $T$ the gross headway). The distribution
$p(v,T)$ has a direct relation to the underlying microscopic states of
the car-following process, which makes it a very interesting object to
study.

\subsection{Headway distributions}

\begin{figure}[ht]
\begin{center}
\includegraphics[width=\linewidth]{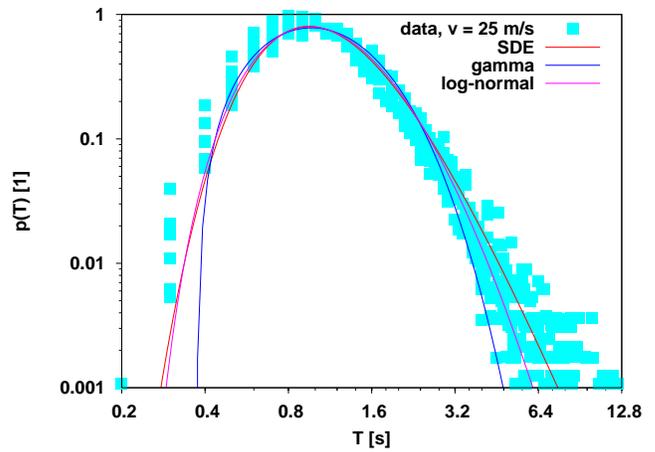}
\caption{Frequency distribution $p(T)$ of the net headway $T$, left lane,
  German autobahn A3 \cite{Schad:data:2002}, speed $v=25$ m/s. The
  empirical distribution is compared to a gamma distribution, to a
  log-normal distribution and to a the distribution
  Eq.~(\ref{eq:theDist}). All functions describing the headway
  distributions have been fitted to the data by a Levenberg-Marquadt
  algorithm.}
\label{fig:pHead}
\end{center}
\end{figure}

One cut at constant speed of the distribution in Fig.~\ref{fig:uFD-3D}
is drawn in Fig.~\ref{fig:pHead}. There, these distribution have been
compared to two of the headway distributions above, and one labelled
with SDE, which will be derived below:
\begin{equation}
p(T) =  {\cal N}  \left ( \frac{1}{T} \right )^{\psi} \exp\left
(-(\psi - 2) \frac{ m_T}{T} \right ) \quad T \ge 0 .
\label{eq:theDist}
\end{equation}
Differently from the other distributions, Eq.~(\ref{eq:theDist}) has a
power-law decay for large headways, with exponent $\psi \approx 4$.

%
%
As could be seen in the plots, the gamma-distribution is not a bad
fit. However, the curve labelled ''SDE'' definitely yields the best
overlap between data and model. Additionally, this distribution needs
just two parameters instead of the three for the gamma-distribution
and the log-normal distribution. The curve SDE is the result of
assuming that the headway of any particular car-driver unit is
controlled by the following stochastic differential equation (SDE)
\begin{equation}
\dot{T} = \alpha ( m_T ( v ) - T )  + D(v) T \xi,
\label{eq:multiNoise}
\end{equation}
where $\langle \xi(t) \xi(t')\rangle = \delta (t-t')$. Here, the
constant $\alpha$ is the inverse relaxation time after which the
variable $T$ returns to $m_T$, and $D$ is the strength
of the noise term that disturbs the convergence to $\langle T
\rangle$. Both parameters $m_T$ and $D$ may depend on
the speed $v$ of the cars. The ansatz above is motivated by the idea
that a driver has a preferred headway $m_T$, which she
is not able to realize exactly and instantaneously. This is, because
human perception and human reactions are usually rather
sloppy. Therefore, a certain time $1/\alpha$ is needed before the
preferred headway is reached, and the process is disturbed by a
stochastic term which models the uncertainty in human perception and
human reaction. It is quite naturally to assume that this uncertainty
becomes smaller when the headway itself is smaller, this is modelled
by the multiplicative noise term $D\, T\, \xi$. A similar process has
been assumed in a completely different context of econophysics. There,
it is used to model the pricing of options \cite{Hull:1987}.

The stationary solution of the corresponding Fokker-Planck equation to
this SDE gives the distribution in Eq.~(\ref{eq:theDist}), with
exponent $\psi = 2+2\alpha/D^2$ and the normalization constant
${\cal N}$:
\[
{\cal N} = m_T \frac{2^{2\alpha/D^2}  \left ( \langle T
\rangle \alpha/D^2 \right )^{2\alpha/D^2}}{\Gamma(2\alpha/D^2)}
\]
%
(Ito calculus, Stratonovich calculus yields a similar result). As can
be seen in Fig.~\ref{fig:pHead}, this function yields the best fit,
especially for larger headways. Here, the power-law behavior is
dominant. Power-laws for different traffic flow variables have
been reported in the past (often in simulations of different traffic
flow models), see
\cite{Nagel:Paczuski,Musha:Higuchi:1976,PW:ZfN:1997,Ben-Naim:1997,Nishinari:2003}
for examples. However, they are mostly related to the distribution of
the speeds, or the lifetime of traffic jams. Rarely, the headway
distributions have been studied.

Therefore, from the analysis of the headway distributions, the
hypothesis may be drawn that the preferred headway of a given driver
is not a constant but is driven by a simple stochastic process. In the
next section, this hypothesis will be discussed again and compared to
alternative formulations that can generate the same distribution.

\subsection{Correlation times}

\begin{figure}[ht]
\begin{center}
\includegraphics[width=\linewidth]{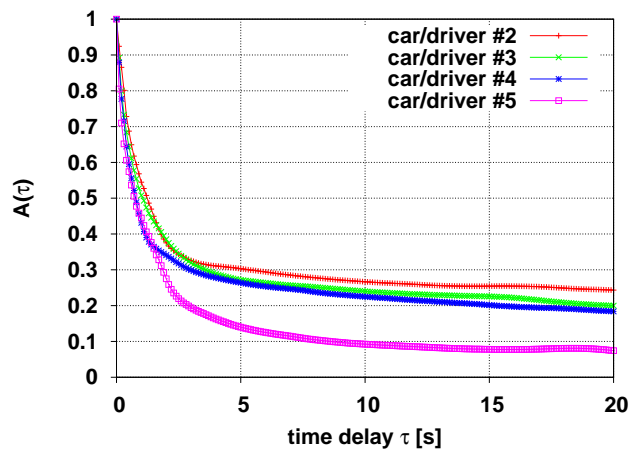}
\caption{Autocorrelation function $A(\tau)$ for the car-following data,
         for four different drivers demonstrating individual
         differences but the same overall behavior.}
\label{fig:autoCor}
\end{center}
\end{figure}

What cannot be deduced from the freeway data is the correlation times
$1/\alpha$ for the headway of a particular car. The headway
correlation function between {\bf different} cars that can be measured
from the freeway data is zero, i.e.\ the headways of different cars
are uncorrelated. The autocorrelation function $A(\tau)$ of the
headway time-series $T(t)$ of one particular car, however, displays a
finite memory. This can be seen in Fig.~\ref{fig:autoCor}, where
$A(\tau)$ for four different cars are presented. They have been
computed from data recorded on a Japanese test track
\cite{Gurusinghe2003,pw2003}. There, the trajectories of ten cars
following a lead car, all equipped with differential GPS, had been
measured. Similar results have been found in other car following data
sets as well; however, the decay times are different. A couple of data
sets describing car-following processes have been made public, see
\cite{CDM}. Of course, if the autocorrelation function follows a
simple exponential delay, namely:
\begin{equation}
A(\tau) \propto \exp(- \alpha \tau),
\end{equation}
then the constant $t_R = 1/\alpha$ is proportional to the relaxation
time of the SDE~(\ref{eq:multiNoise}). It is not equal to the
relaxation time, because the stochastic process driving the headway is
filtered through the set of differential equations that describe the
car-following dynamics under the assumption that $T$ is constant. The
actual result seems to be more complicated, it is consistent with two
superposed exponential decay curves $A(\tau) = w
\exp(-\tau/\tau_{\text{fast}}) + (1-w) \exp(-\tau/\tau_{\text{slow}})$, 
where the fast decay happens within $\tau_{\text{fast}} = 2\ldots6$ s,
while the slower decay took about a factor of ten longer. In general,
the $\tau_i$ are shorter in dense traffic, for a more detailed
analysis the available database is still too narrow.

The two components of the autocorrelation function can be traced back
(not shown) to the autocorrelation of the gaps (long-lived component)
and of the speed-differences (short-lived), respectively. (Both enter
in the computation of $T = g/v$.)

\subsection{Dependence on speed}

The parameters $m_T$ and $D^2/\alpha$ may depend on speed, too. For
large speeds, $m_T$ diverges. This stems from the bigger distances
between the freely moving cars. Not as simple to understand is the
divergence for the small speeds, later on this will be analyzed more
detailed.
\begin{figure}[ht]
\begin{center}
\includegraphics[width=\linewidth]{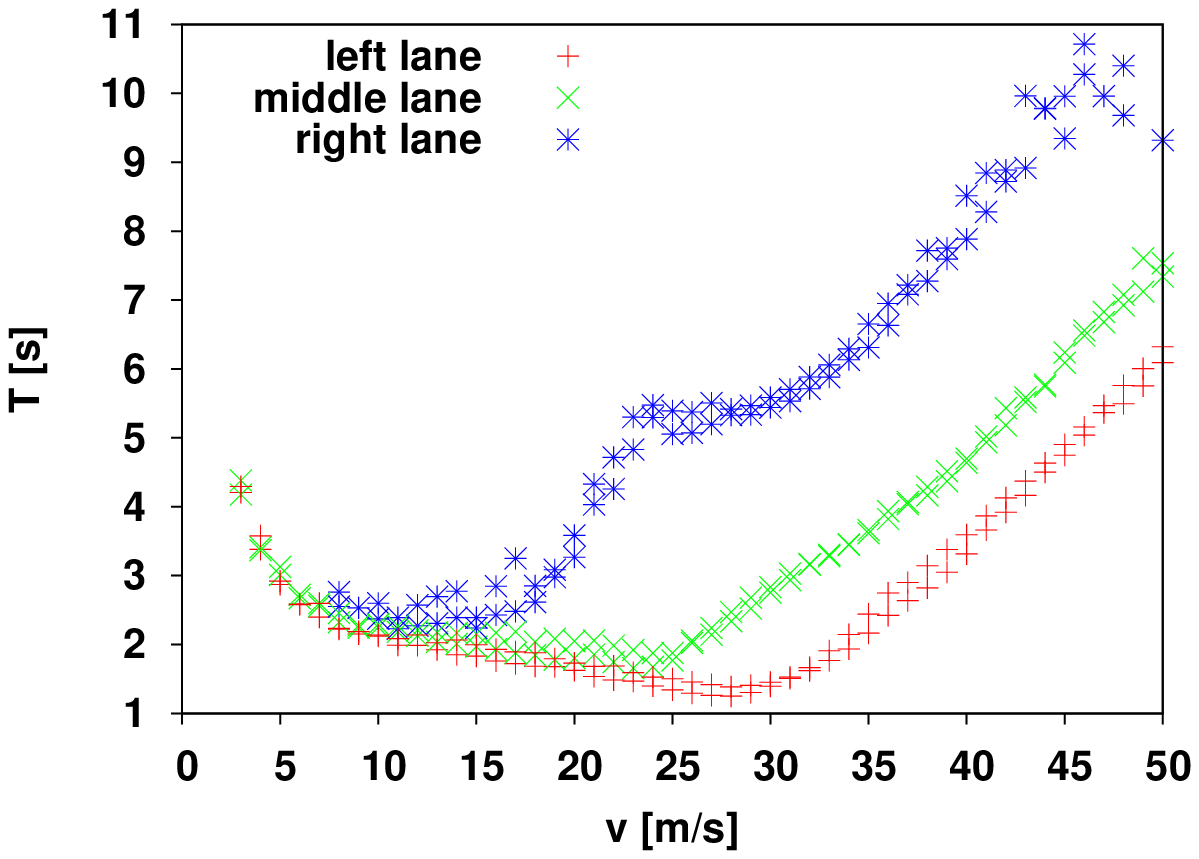}\\
\includegraphics[width=\linewidth]{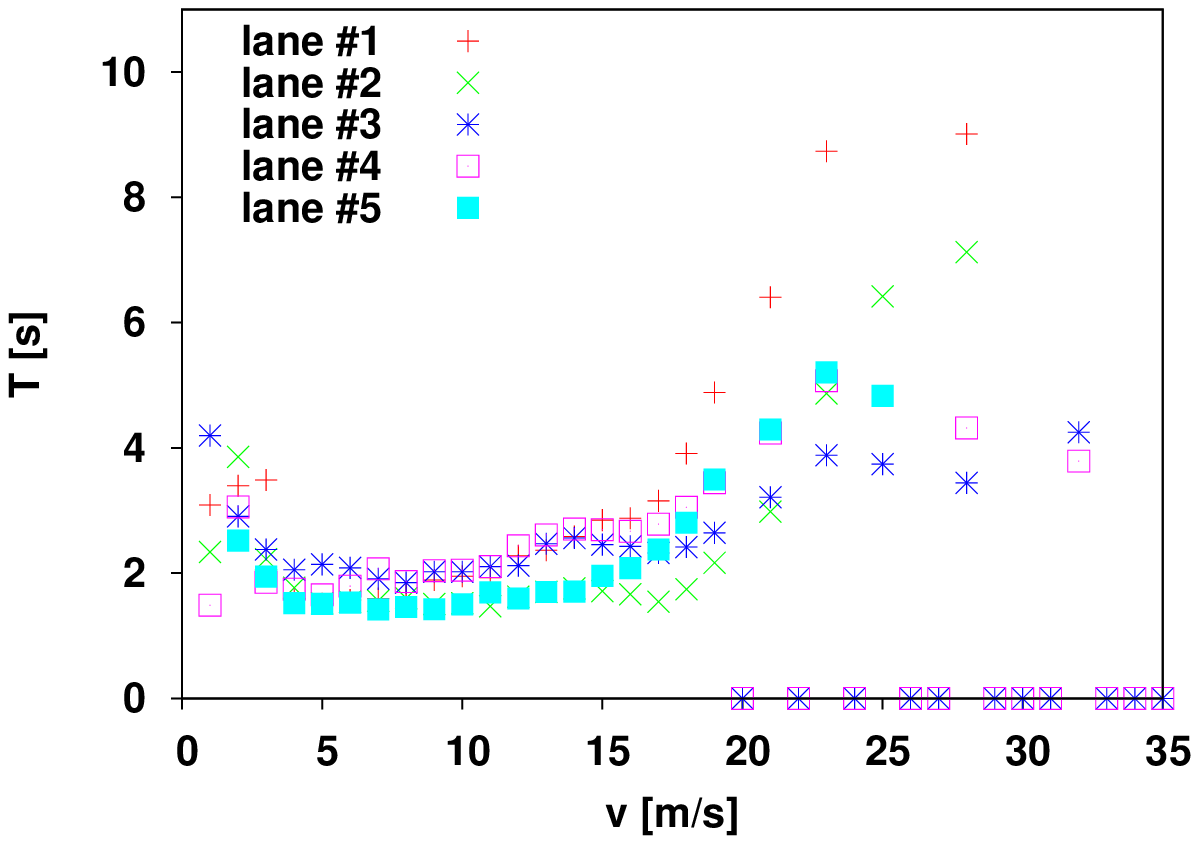}
\caption{Mean value of headway $T$ as function of 
  the speed $v$. The upper plot is for the German freeway A3, while
  the lower plot is for the California freeway I-880.}
\label{fig:uFD}
\end{center}
\end{figure}
Additionally, there is an intermediate region, where the mean headway
depends in a non-trivial manner on the speed of the cars, see
Fig.~\ref{fig:uFD}. Between two values $v_1$ and $v_2$ which depend on
the circumstances (especially the speed limit), the dependence of
$m_T$ on speed becomes a decreasing function of speed. This empirical
finding has been mentioned already in \cite{IDMM}. Note, that this is
a real effect, since there is no physical reason for the drivers to
relax, i.e.\ to drive with larger headways \footnote{Obviously, a
  psychological reason might be the fact that they cannot go faster in
  congested traffic, so there is no need to drive with very short
  headways. However, when physicists use this kind of arguments, great
  care should be taken.}.  On the contrary, anybody would be better
off if car-drivers try to keep those short headways, since it would
increase the throughput on a highway.

Although the most direct explanation for this dependence of $\langle T
\rangle$ on speed $v$ is that of more relaxed driving, an alternative
interpretation is possible. Assuming that drivers have different
driving attitudes, the increase in headway may be related to a change
in the driver population. Borrowing from
\cite{daganzo:mlane-I,daganzo:mlane-II} the self-explanatory terms
rabbits and slugs, it may be assumed that for the smaller speeds more
slugs populate the left lane, increasing the mean headway.

Interestingly, this seems to be true only under certain
circumstances. It has been found in the data set from the left lane of
the German freeway A3, when looking on the other lanes, the dependence
is much less clear. Furthermore, drivers on American freeways seem to
behave differently. In \cite{Banks:2003} it is demonstrated, that the
mean headway is independent of speed. An analysis of data measured in
the FSP-project (data and description of this project can be found at
\cite{CDM}, too) and from another project on the California freeways
I-880 and I-80, respectively, support this result. Using once more the
terms rabbits and slugs above, on American freeways the different lanes
are more homogeneous, since there is no reason for the slugs to stay
on the right lane(s).

\begin{figure}[ht]
\begin{center}
\includegraphics[width=\linewidth]{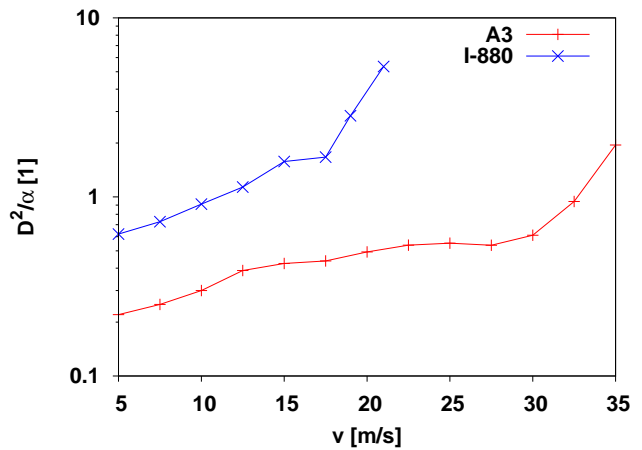}
\caption{Noise strength $D^2/\alpha$ as function of speed $v$. For 
         smaller speeds, the noise is smaller. The numbers for
         $D^2/\alpha$ have been obtained by fitting each $p_v(T)$
         histogram with a Levenberg Marquadt algorithm. The I-880
         site had a speed limit of 55 mph (24.6 m/s), the data are restricted 
         to the rush-hours, so larger speeds than 22 m/s do not contain
         enough statistical weight to allow for a similar analysis.}
\label{fig:noise-vs-speed}
\end{center}
\end{figure}

That is not all. As can be suspected, the noise strength, too, is a
function of speed, with $D^2/\alpha$ decreasing as speed
decreases. This can be seen in Fig.~\ref{fig:noise-vs-speed}.
While the overall dependence is the same for the US-example as well as
for the German A3 data, the width of the headway distribution in the
American data is bigger by about a factor of almost three.

\subsection{Slow--to--start behavior}

\begin{figure}[ht]
\begin{center}
\includegraphics[width=\linewidth]{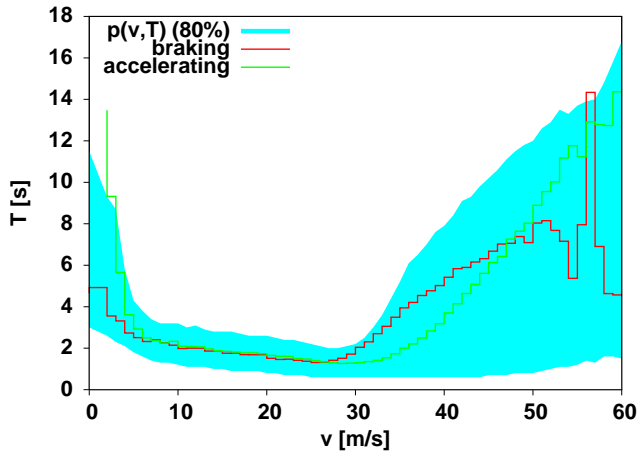}
\caption{Slow-to-start behavior. Plotted is headway $T$ versus speed $v$. 
         The filled area represents the 
         center 80\% of the mass of the frequency distribution
         $p(v,T)$. The line labelled acceleration is from cars where
         $v_i > v_{i-1}+v_c$ holds, while the braking line is from
         cars with $v_i < v_{i-1} - v_c $. The threshold $v_c$ has
         been set to $v_c = 0.5$ m/s.}
\label{fig:slow2start}
\end{center}
\end{figure}

While the creation mechanism of jams is still controversial, the fact
that jams are stable can be understood much more easily. The simplest
idea that causes stable jams is what has been called quite tellingly a
slow-to-start mechanism \cite{Sch:Sch:slow-to-start}. It is assumed,
that a car leaving a jam needs considerably more time-headway than
what is needed in maximum flow conditions. Already in
\cite{Newell1962} such a mechanism is proposed, to explain the
jam-waves observed in the Holland tunnel in New York. However, the
empirical support reported so far is weak. E.g.\ the model proposed in
\cite{Newell1962} assumes, that the fundamental diagram in the
congested region consists of two branches, one for accelerating
and the other one for decelerating traffic. Usually, the scatter
in the empirical data is so big, that the two branches are
completely hidden by the scatter, if they exist at all (see
Fig.~\ref{fig:slow2start}). Again, the single-car data analyzed
here help to explore this putative mechanism in greater detail.
This is displayed in Fig.~\ref{fig:slow2start}. There, the
headway velocity data have been averaged for cars where acceleration
or deceleration can be assumed. Since acceleration or deceleration is
not measured in these data, it is assumed that a car is accelerating
when $v_i > v_{i-1}+v_c$ holds. Vice versa, if $v_i < v_{i-1}-v_c$
holds, it is decelerating. Definitely, a more thorough analysis had to
wait for trajectory data. Currently, there are a couple of projects
running world-wide to collect those data, the one that is probably
most advanced is \cite{NGSIM-2004}.  Nevertheless, two interesting
observations can be made already with the single car data available
for this study. First, there is indeed a slow-to-start regime for
speeds smaller than 5 m/s. By comparing with a simulation where this
feature has been explicitly put in (see next section for details), it
could be stated that the increase in headway is not restricted to
accelerating cars. The simulation data coincide with the empirical
data only if the assumption is stated, that for small speeds (below 5
m/s) car drivers increase their time headway considerably.

The second interesting feature in Fig.~\ref{fig:slow2start} is that
for larger speeds there is a regime where the decelerating cars
(instead of the accelerating cars) have the larger time headway. The
latter feature is hard to understood right now, it certainly needs
more work to explain and is most likely related to multi-lane
phenomena.

\subsection{Regimes of traffic flow}\label{sect:regimes}

Figure~\ref{fig:slow2start} already demonstrates, that there are
different regimes of traffic flow which may be discernable on the
basis of the mean headway for accelerating and decelerating cars. The
different regimes can be demonstrated more clearly by analyzing the
standard deviation of the speed differences between following cars:
\begin{equation}
\Delta^2(v) = \frac{1}{N}\sum_{i=1}^N (v_i - v_{i-1})^2 - 
\left ( \frac{1}{N}\sum_{i=1}^N (v_i - v_{i-1}) \right)^2.
\end{equation}
The speed $v$ used for reference on the left hand side of the equation
is $v_{i-1}$, i.e.\ the expression above measures the width of the
distribution of speed differences, which in fact depends on the speed
itself: for large speeds, corresponding to free flow conditions, speed
differences can be fairly large, while they become small under
congested conditions (smaller speed).

\begin{figure}[ht]
\begin{center}
\includegraphics[width=\linewidth]{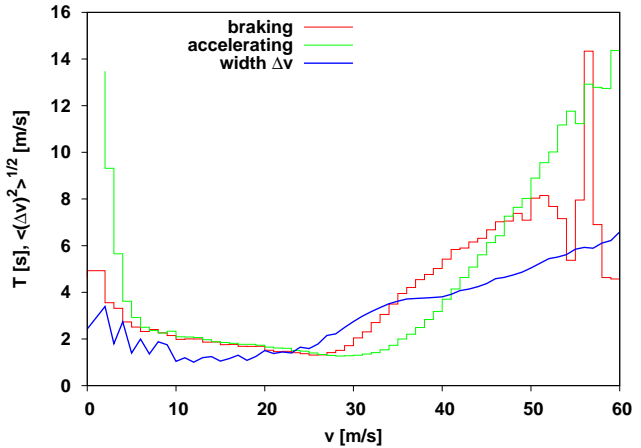}
\caption{The function $\Delta(v)$ for data from the German freeway 
         A3, together with the headway data determined as in
         Fig.~\ref{fig:slow2start}, plotted here for reference. }
\label{fig:sigma:v}
\end{center}
\end{figure}

The function $\Delta(v)$ which is shown in Fig.~\ref{fig:sigma:v} displays a
fairly complicated behavior.
At least, three different regimes can be separated: a small speed
regime, an regime of intermediate speeds where $\Delta_v(v)$ is small
an roughly constant, and a high-speed regime, where $\Delta_v(v)$
increases with speed. What all this means becomes clearer by looking
at $\Delta_v$ as function of the fundamental diagram. The region with
comparably small $\Delta_v$ is found in the middle of the fundamental
diagram, and this is more or less the region that is usually
identified with synchronized flow (see \cite{KernerPRE2002} for a
good summary of all the previous work on synchronized flow).
\begin{figure}[ht]
\begin{center}
\includegraphics[width=\linewidth]{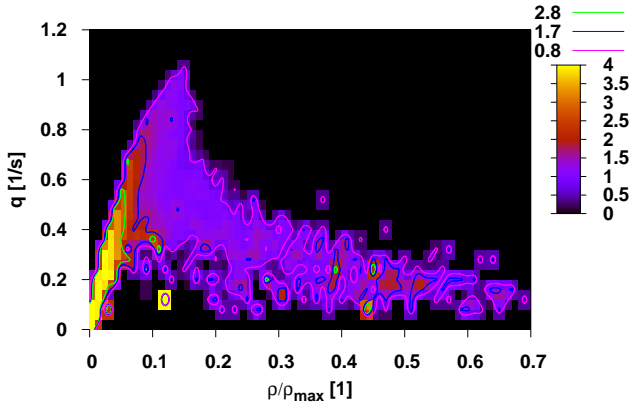}
\caption{The standard deviation of speed-differences, $\Delta_v$, as 
         function of occupancy $\rho/\rho_{\text{max}}$ and flow $q$,
         again for data from the left lane of the German freeway A3. Note the plateau
         of constant $\sigma_v$ (the contour lines labelled with 0.8 and 1.7
         m/s).}
\label{fig:FD-sigma:v}
\end{center}
\end{figure}
This means, that the variable $\Delta_v$ as defined above gives a
fairly well statistically meaningful characterization of the states of
traffic flow, which are associated with synchronized flow. However,
it could be stated that the region of small $\Delta_v$ extends to
the left, into the high-flow regime. The high-flow regime is assumed
to be different from the synchronized flow regime, however the
Fig.~\ref{fig:FD-sigma:v} here suggests that even the high flow state
is just a synchronized state of traffic flow. To make clear that this
is different from the view in \cite{KernerPRE2002}, this entire region
will be called homogeneous flow of interacting cars in the following.

A similar conclusion has been drawn already in \cite{Helbing:block}.   
With respect to the interaction, it is
quite naturally to recognize high flow states as similar to the other
states where cars are interacting heavily: in both cases, the driving
behavior is very concentrated and attentive of the surrounding cars,
otherwise those high-flow states could not be sustained.

To summarize, three different regimes of traffic flow may be discerned
clearly: the low demand regime, where the headway distribution is
Poissonian, the interaction regime where headways are power-law
distributed, and a jammed region, where headways probably are
something different. The interaction-dominated regime is the one of
the small $\Delta_v$.

This issue will be discussed once more in the light of the discussion
about the model to be described next.

\section{Consequences}\label{sec:YAM}

So far, empirical results have been presented. In this section, the
consequences for microscopic models will be discussed. For the
macroscopic models, similar ideas may be followed.

Before explicitly demonstrating how the empirical results can be
implemented into a certain microscopic model of traffic flow, the
basic assumption of this work should be discussed more detailed. It
has been assumed, that the preferred headway of a driver follows a
simple stochastic process defined by Eq.~(\ref{eq:multiNoise}). There
are (at least) two additional hypotheses that lead to the same or a
similar distribution. The first one is to assume that the $T_i$ of a
driver is constant \cite{Cassidy:memory:1998} and that the $T_i$ of
all the drivers are distributed according to the distribution
Eq.~(\ref{eq:theDist}).  The second hypothesis assumes a stochastic
process for $T$ without memory. Both hypotheses can be rejected: the
first alternative because the empirical data from the car-following
experiments demonstrate that $T$ is not constant for a particular
driver. The second alternative can be ruled out by comparing
simulation results of the model defined next with different stochastic
processes for the headway. This results in a headway distribution that
is definitely different for the white-noise assumption than it is for
a time headway driven by Eq.~(\ref{eq:multiNoise}). (This does not rule 
out that the $m_T$ of the drivers are distributed, too.)

\subsection{Models}

As a simple example, the model introduced in \cite{Krauss:Metastable}
will be extended with the empirical results
above. Obviously, these results can be transferred to most of the
known models, provided they have something like a preferred
distance. While this will lead to the correct headway distribution,
the macroscopic behavior of different models might be different. The
model \cite{Krauss:Metastable}, and a very similar model \cite{Gipps:following} can
be derived from a safety condition, namely:
\begin{equation}
d(v_i) + v_i \, T_i \le d(v_{i-1}) + g_i
\label{eq:SK-safety}
\end{equation}
where the $d(v_i) = v_i^2/(2b)$ are the braking distances of the
following (index $i$) and the leading car (index $i-1$),
respectively. The deceleration $b$ is understood as a comfortable
braking deceleration, not the maximally possible one. The constant
$T_i$ is the preferred headway of the driver, $g_i^\star=v_i T_i$. The
equation above can be solved to yield the safe speed:
\begin{equation}
v^{(i)}_{\text{safe}} = -b T_i + \sqrt{(b T_i)^2 + v_{i-1}^2 + 2bg_i} .
\label{eq:SK-vsafe}
\end{equation}
From this, the final model in \cite{Krauss:Metastable} is constructed by assuming 
an update equation of the form (denoting time as $t$ and the time step size by
$h$):
\[
v_i(t+h) = \min \{v_i(t) + a_i h, v_{\text{safe}}, v_{\text{max}} \} ,
\]
where the constants $a_{\text{max}}, v_{\text{max}}$ are parameters
(maximum acceleration and maximum speed, respectively) of the
model. (The complete model additionally contains a noise term.)
However, this equation describes the speeds and not the acceleration,
the speed adapts to the safe velocity almost instantaneously (within
one time-step $h$). To define an equation for the acceleration, the
model can be formulated similar to the optimal velocity
models~\cite{Newell1962,Bando:etc:pre}:
\begin{equation}
a_i = (v^{(i)}_{\text{safe}} - v_i)/T_{\text{acc}}
\label{eq:yam-asafe}
\end{equation}
This describes the relaxation of the current speed towards the safe
speed, with a time constant $T_{\text{acc}}$. This time constant is
basically the autocorrelation time of the resulting acceleration time
series. Extracting this number from an acceleration time-series
typically gives values between one and three seconds for this
constant.

Equations (\ref{eq:SK-vsafe}),(\ref{eq:yam-asafe}) define the
deterministic part of the model; it does not contain an explicit white
noise term that acts on the acceleration. This (white) acceleration
noise term that is often used in physical models of traffic flow is
clearly unrealistic, and the stochastic process defined in
Eq.~(\ref{eq:multiNoise}) is an alternative. Together with a discrete
version of Eq.~(\ref{eq:multiNoise}), the model is fully specified:
\begin{eqnarray}
x_i(t+h) & = & x_i(t) + v_i(t) \, h
\label{eq:yam-pos} \\
v_i(t+h) & = & v_i(t) + a_i(t) \, h \label{eq:yam-speed} \\
T_i(t+h) & = & T_i(t) + \alpha h  (m_T - T_i) + \nonumber \\
 & & \sqrt{h} D T_i \xi_i,
\label{eq:yam-noise}
\end{eqnarray}
with $a_i(t)$ as given in Eq.~(\ref{eq:yam-asafe}). Here, $\xi_i(t)$
is a random number in $[-1,1]$. Formally, Gaussian random numbers
should be used here. To get a fast implementation, this can be omitted
\cite{Honerkamp:Book}. Of course, to describe freeway traffic
realistically, $m_T(v)$, $D(v)$, and the parameters needed by the
model have to be provided. They can be obtained from the empirical
data such as Fig.~\ref{fig:uFD} or by directly fitting this model to
traffic flow data, see \cite{Brockfeld2003a,Brockfeld2004a} for
examples. The numbers found from the freeway data are $\alpha = 1/2 \ldots 1/5$
s$^{-1}$, $D=0.5$, $T_{\text{acc}} = 1.0$ s.

The model can be made crash free by enforcing in each update step $v_i
\le v_{\text{safe}}$.

\begin{figure}[ht]
\begin{center}
\includegraphics[width=\linewidth]{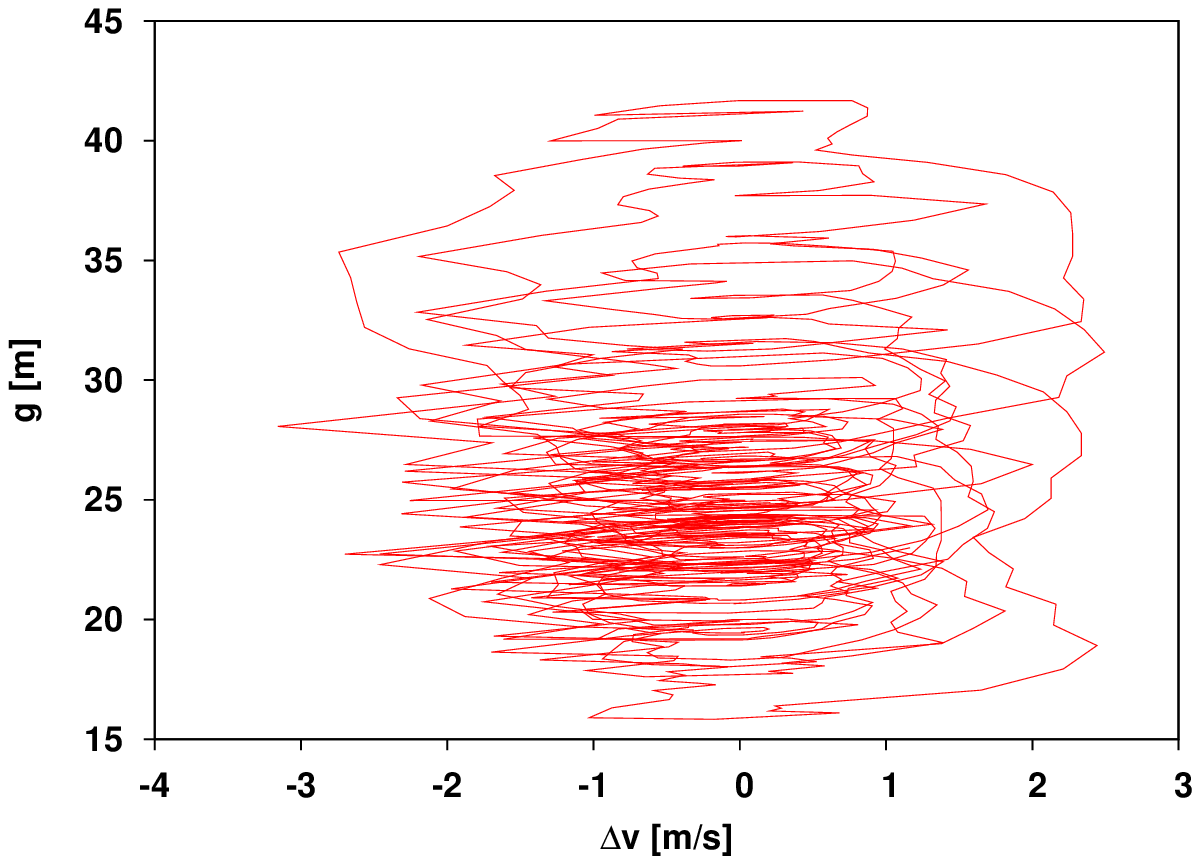}\\
\includegraphics[width=\linewidth]{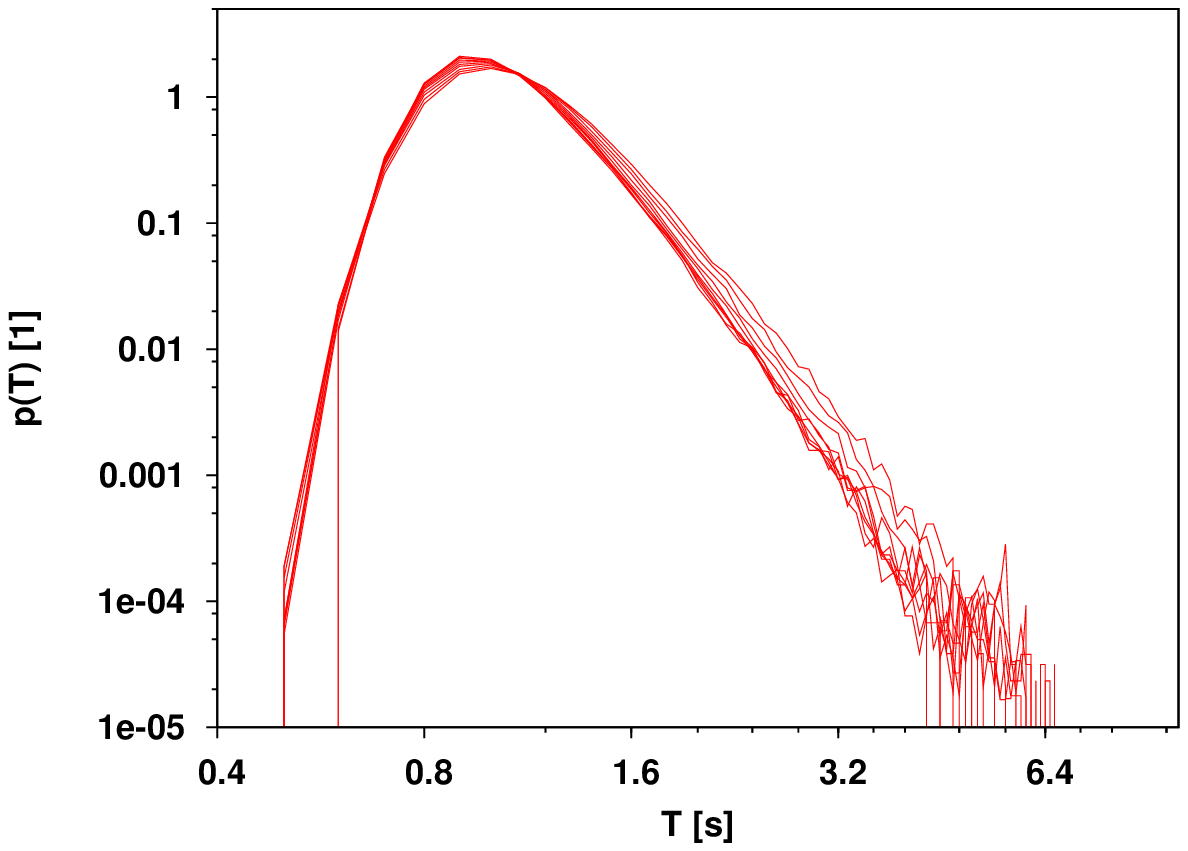}
\caption{Oscillations in speed difference $\Delta v$, distance $g$
         space (above) and the headway distribution (below) for a simulation
         run with the model defined in Eqs~\protect{(\ref{eq:yam-pos})
         -- (\ref{eq:yam-noise})}. Parameters are $D=0.5$,
         $\alpha=1/5$ s$^{-1}$, $h= 0.4$ s, $m_T = 1$
         s. The driving parameters were set to $a_{\text{max}} =1.5$
         m/s$^2$, $b=2$ m/s$^2$, and $v_{\text{max}}=31$ m/s. A
         simulation with $n=200$ cars on a closed loop was run for
         10000 s before data were sampled for another 10000 s.}
\label{fig:uSim}
\end{center}
\end{figure}

A more detailed analysis of this new model will be performed
elsewhere. Here, the behavior of the model with respect to the
question of the traffic flow phases will be discussed. However, even
microscopically the model is in line with two important empirical
observations. First, it yields the noisy oscillations in
speed-difference headway space, see Fig.~\ref{fig:uSim}. Secondly, the
time headway distribution is very similar to the empirical headway
distributions.
%

\subsection{Macroscopic behavior}

\begin{figure}[ht]
\begin{center}
\includegraphics[width=\linewidth]{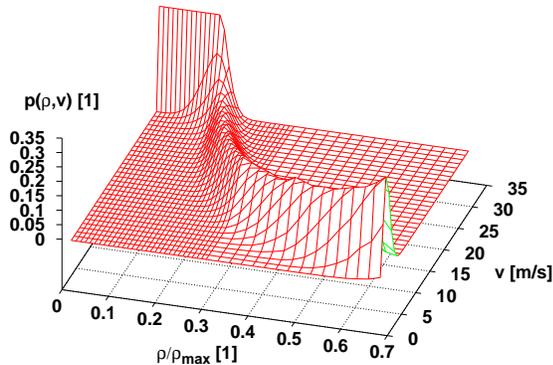}
\caption{Fundamental diagram of the model Eqs~\protect{(\ref{eq:yam-pos})
         -- (\ref{eq:yam-noise})}.  Parameters and simulation set-up
         are the same as in Fig.~\ref{fig:uSim}. }
\label{fig:yam-uFD}
\end{center}
\end{figure}

The model as defined above has only one solution as function of the
density: a homogeneous state, with a width in the time headway
distribution given by $D^2/\alpha$ of the SDE above. This can be seen
in the fundamental diagram, see Fig.~\ref{fig:yam-uFD}.
The mean value $\langle v \rangle$ of this distribution is, of course,
a function of density $\rho = 1/(g + \ell)$, where $\ell$ is the
length of a car, and $\rho_{\text{max}}=1/\ell$ is the maximum
density. It can be obtained approximately analytically by setting
$a_i=0$ in Eq.~(\ref{eq:yam-asafe}), integrating the resulting
expression over the distribution $p(T)$ of and solving for
$v(g)$. This yields:
\begin{equation}
\langle v(\rho) \rangle \approx \left \lbrace
\begin{array}{ll}
v_{\text{max}} & \mbox{if} \quad \rho < \rho_c \\
\frac{1}{m_T} \left ( 1/\rho - \ell \right) &
\mbox{else}
\end{array}
\right .
\end{equation}
This formula fits quite well the simulated fundamental diagram,
however a slightly larger value of $m_T$ is needed to
obtain an exact fit.

However, the model is not complete, because it lacks a mechanism that
stabilizes a jam once it is created. As pointed out above, the
empirical data analyzed above seem to support the idea of a
slow-to-start mechanism. There are different means to implement such a
mechanism. The one chosen here is to alter the relation for
$v_{\text{safe}}$ by making $v_{\text{safe}}$ smaller for small
speeds. This is done in Eq.~(\ref{eq:SK-safety}) to arrive at a
modified $\tilde{v}_{\text{safe}}$:
\begin{equation}
\tilde{v}^{(i)}_{\text{safe}} =
\left \lbrace \begin{array}{ll}
v_{\text{safe}} - \epsilon (v_i - v_c)  & \mbox{if} \quad v < v_c \\
 v_{\text{safe}} & \mbox{else}
\end{array} \right .
\label{eq:SK-s2s}
\end{equation}
Of course, $\tilde{v}_{\text{safe}} > 0$ must be demanded explicitly.

This changes the behavior of the model. It becomes bistable in some
regions of the control parameter space $\rho,D$. The
corresponding phase diagram is displayed in
\begin{figure}[ht]
\begin{center}
\includegraphics[width=\linewidth]{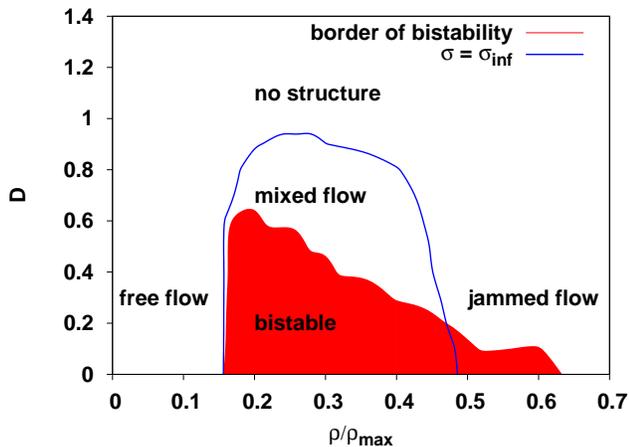}
\caption{Phase diagram of the model Eqs~\protect{(\ref{eq:yam-pos})
         -- (\ref{eq:yam-noise})}. Parameters are as in
         Fig.~\ref{fig:uSim}, additionally a slow-to-start mechanism
         as in Eq.\ref{eq:SK-s2s} has been adopted. Simulation has been 
         run for 40000 s, with 500 cars for each value of $\rho$ and $D$. 
         See text for details on how the separating lines have been obtained.}
\label{fig:phaseDiag}
\end{center}
\end{figure}
%
Fig.~\ref{fig:phaseDiag}. To demonstrate bistability, the system has
been started either in a jammed state or in a homogeneous state. After 
running for 40000 s, the probability distributions $p^{\text{hom}}(v)$ and 
$p^{\text{jam}}(v)$ are being analyzed. For bistable parameter values, 
there should be a difference between the two distributions, which can be 
measured by computing $\eta = \int p^{\text{hom}}(v) - p^{\text{jam}}(v) dv$. 
This is a number between zero and one, with $\eta=0$ indicating equal outcomes 
of the simulation. The shaded area in the plot above is the area where $\eta>0.1$. 

The different areas free flow, jammed flow, mixed flow, and unstructured have been 
assigned on the basis of the standard deviation of the speed $\sigma(\rho,D)$ 
 which can be computed  from $p(v)$, too. The whole approach is very similar to 
\cite{Nagel:breakdown}. Let's start with the unstructured state. When increasing the noise 
$D$ of the headway distribution, the speed distribution must finally approach the constant 
distribution in $ [ 0, v_{\text{max}} ] $, with standard deviation $\sigma =: \sigma_{\infty} = v_{\text{max}}/\sqrt{12}$. The simulation results confirm this. This state may be
called, in analogy with equilibrium thermodynamics, the high
temperature state.

For smaller values of $D$, the system can be in three different states: a free flow state, where all cars
move with approximately the free speed, the jammed state where all cars move with 
small speeds near zero, and a mixed state where the system consists of a mixture 
of freely moving and jammed cars. This picture of nucleation is quite familiar from
the theory of (equilibrium) phase transitions. Additionally, the system may be in the 
homogeneous
state, where all cars move with the same speed, which is smaller than the free flow 
speed. For small $D$, the standard deviation of speed therefore is very small for small
densities, since $p(v)$ is close to a delta-function there. The same is true for very
large densities. In between, the system may consist of a mixture of jammed and free driving
cars, in this case the standard deviation of speed becomes maximal, because $p(v)$ consists
of two delta-peaks located at $v=0$ and $v = v_{\text{max}}$, respectively. In this case, 
$\sigma = v_{\text{max}}/2$. For moderate values of $D$, the peaks are wider, 
therefore $\sigma$ descreases until it finally reaches  $\sigma_{\infty}$. 

Therefore, the line where $\sigma = \sigma_{\infty}$ can be used to discern these
states, and that is what is drawn in Fig.~\ref{fig:phaseDiag}.

The homogeneous state, which does not fit into the picture borrowed from thermodynamics,
is stable only for small $\rho$ and small $D$, for larger values it decays. 
With respect to phenomenology of traffic flow, it is important that there is a
region in this phase diagram the homogeneous state co-exist with the mixed state, at least to
the time resolution applied here (each simulation for each of the
data-points $\rho,D$ has been run for 40000 s). What happens depends
on the initial conditions, or, in the case of an open system, on the
boundary conditions applied. Such a bistability is a very attractive
feature, since it gives an idea why different states of traffic may be
observed in cases where anything else is identical. This gets
additional back-up with the observation, that the empirical data leads
to a value of $D=0.5$, which is in the bistable region (depending, of
course, on $\rho$).

\subsection{Comparison with empirical data}

\begin{figure}[ht]
\begin{center}
\includegraphics[width=\linewidth]{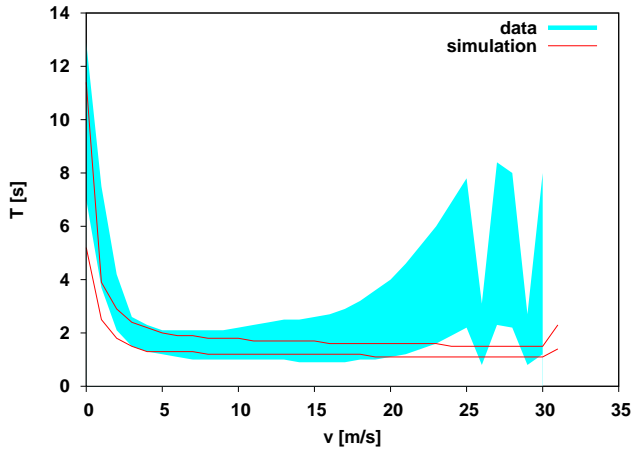}
\caption{The 25th-percentile and the 75th-percentile lines of the 
             empirical probability distribution $p(v,T)$ compared to simulation 
             data. For the simulation, the same set-up as in Fig.~\ref{fig:uSim} has 
             been adopted. The simulation data are from a scan of the fundamental diagram,
             i.e.\ a simulation has been run for different densities.}
\label{fig:compData}
\end{center}
\end{figure}

The model compares qualitatively very well with real data. However, it
is still not a good model in the sense of comparing well even
quantitatively with empirical traffic flow data. However, on the level
of macroscopic measures like the headway distribution $p(v,T)$, the
comparison is quite successful, as Fig.~\ref{fig:compData}
demonstrates.
There, the microscopic fundamental diagram has been compared with the
empirical one. Although such comparisons are of limited value, it
demonstrates that at least the fluctuations of traffic flow are well
described by the model introduced here.

To make this comparison more quantitative, new features must
added. These include, but are not restricted to, good ideas how
crash-freeness is achieved by human drivers, anticipation to the
behavior of other cars in front, more details on how the acceleration
is chosen by a driver, issues related to driver inhomogeneity, and
multi-lane modelling, see \cite{pw2003} for more details.

\section{Conclusions}

\subsection{Synchronized flow}

As discussed already in section~\ref{sect:regimes}, the states of
traffic flow where the interaction between the cars is important have
a relation to the synchronized traffic flow. We will not enter the
discussion about the truly complicated phenomenology of synchronized
flow, see \cite{KernerPRE2002} for details on that. However, a
synthesis is tried that unifies the simulation results with the
empirical results, and further more, may bring together the different
views about synchronized flow put forward so far (see
\cite{KernerPRE2002,daganzo:critique:sync,SchreckiPRL2004}).

The idea here, which is motivated by the surprising stability of
the homogeneous flow of the traffic flow model above (and in fact
of most traffic flow models, to make them truly unstable is not a
simple task), is as follows. There is a large region of traffic
flow states where the homogeneous state as function of density is
stable. Definitely, homogeneous does not mean that the relevant
variables like flow, density, and speed are distributed sharply.
In fact, this work here is about a clean description of the
fluctuations in the headways of the cars, so homogeneous means a
distribution of a certain width. Only for very large densities, or
for truly strong external perturbations, there is in fact an
instability where breakdowns could occur. These breakdowns are
probably related to the pinch effect described in \cite{KernerPRE2002}.

So, when such a system is regarded on a closed loop (no boundary
effects are present), just three different states might be possible: a
free flow state with a Poissonian headway distribution, a homogeneous
state with a power-law headway distribution, and a jammed region. In a
certain range of densities and noise amplitudes $D$ the jammed state
and the homogeneous state are both stable, i.e.\ different initial
conditions lead to different final states. Whether or not there is a
phase transition in the sense of non-equilibrium physics between these
different states is right know not clear. Even if the system manages
to stay at the homogeneous state while increasing the density, there
is a transition where the Poissonian distribution is translated into
the power law distribution. Although there is a really nice symmetry
between these two states (the gamma distribution for $T$ transforms
into the power-law distribution under the transformation $T \to 1/T$),
it is not clear whether this is in fact a phase transition. But even
if it is, it is one of second or even higher order, because nothing
jumps here: it is simply the number of cars that drive freely which is
reduced until finally all cars are following its lead car, without any
chance to overtake.

In the case of an open system, more can happen. Any capacity drop
downstream caused by an incident, or a change of inflow at an on-ramp,
causes what have been called the back-of-the-queue states, and the
transition from apparently free flow into this state is a
discontinuous, first order phase transition.  (Speed jumps from a
large to an intermediate value.) This congested state with reduced speed 
is exactly what has been called homogeneous flow of interacting cars
above. In fact, some of the data analyzed in this
contribution (data from the I-80 in California) are from a site where
a toll station exists downstream which causes the daily break-down in
this area.

This is in line with theoretical considerations about very simple open
traffic flow systems. There, it could be demonstrated explicitly, that
the transition from a free flow state to the congested flow dominated
by the capacity constraints downstream, is in fact a first-order
transition
\cite{Schuetz1999,SchuetzInDomb,Cheybani:2001a,Cheybani:2001b,Helbing:5phases,Namazi2002,Rajewsky,Appert2001}. So, what may look like a first order transition of 
the bulk system might in fact be a transition caused by the boundary.


\subsection{More general remarks}

There is one final question to be answered: why is the headway driven
by such a process? Part of the answer might be what has been proposed
recently as the human driver model \cite{Helbing:HDM}. There, it is
assumed that a driver is simply not capable of estimating the distance
and the speed difference to the car in front timely and
accurately. This is certainly true, and the autocorrelation results
presented in this paper lend empirical support to such an
idea. However, it introduces two things that are truly hard to
measure: how wrong do humans estimate these two numbers, and how do
they modify their outdated measurements. It should be simple to
demonstrate, that in the context of the model above, this assumption
leads to a similar result for the time headway distribution, however,
preliminary simulation results demonstrate that the resulting model
displays a different behavior, so this needs a completely fresh
approach.  Unfortunately, \cite{Helbing:HDM} does not provide any
details on the resulting frequency distributions. Proposing just one
stochastic process for the time headway alone is more in line with
Occam's razor asking for the smallest number of assumptions
necessary. And, in this case, parts of this assumed stochastic process
can be measured almost directly. For example, except for the
relaxation time of the SDE Eq.~(\ref{eq:multiNoise}), all the
parameters that enter into the model proposed here can be measured
explicitly. Nevertheless, assuming a headway driven by such a
stochastic process is for sure not a complete description of reality:
to do that, much more work on the psychology of human driving needs to
be done. Nevertheless, the results achieved here can be summarized as
follows. The model proposed here is a model in the best sense: as
minimal as possible (hopefully), a little bit abstract, and
(certainly) a little bit wrong.

\section*{Acknowledgement}

I would like to thank Reinhard Mahnke for inviting me to a small but
beautiful conference, where I learned about the stochastic process
which drives $T$.  Regarding the data: I am deeply indebted to
T.~Nakatsuji and his Hokkaido group for sharing their data. Those data
provided really valuable insights. Other donations of data came from
the Duisburg group of Michael Schreckenberg, which I acknowledge here
as well. Furthermore, discussions with Carlos Daganzo, Ihor
Lubashevsky, Kai Nagel, Ralf Remer, Andreas Schadschneider, and
Kai-Uwe Thiessenhusen helped to clarify the ideas presented here.

\bibliographystyle{plain}
\bibliography{pw}

\end{document}